\documentclass[intlimits,twoside,a4paper]{article}

\usepackage[cp1251]{inputenc} %

\usepackage{cmpj3} 

\issue{2023}{26}{4}{43803}
\doinumber{10.5488/CMP.26.43803}

\title[Structural transition in biomolecules by perturbation]%
{Structural transition induced by a local chemical/mechanical perturbation in biomolecules} %
\author[F. Hirata]{F. Hirata\orcid{0000-0002-9430-0754}\thanks{hirata@ims.ac.jp}} 
\address{National Institutes of Natural Sciences, Institute for Molecular Science, Myodaiji, Okazaki, Aichi 444-8585, Japan} %


\begin{document}

\maketitle

\begin{abstract} 
Structural transition induced by a local conformational change in biomolecules is formulated based on the generalized Langevin theory for the structural fluctuation of a molecule in solution, and the linear response theory, derived by Kim and Hirata in 2012. A chemical/mechanical change introduced at a moiety of biomolecules, such as an amino acid substitution or a structural change of a chromophore by the photo-excitation, is considered as a perturbation, and the rest of the protein as the reference system.
The linear-response equation consists of two parts: a mechanical/chemical perturbation introduced at the moiety, and the variance-covariance matrix of the reference system that works as a response function. The physical meaning of the equation is transparent: the force exerted by atoms in the moiety induces the displacement in an atom of protein, which propagates through the variance-covariance matrix to cause a global conformational change in the molecule.
A few examples of possible application of the theory, including those in industry, are suggested.
%
%
\printkeywords %
\end{abstract}

\section{Introduction}

It is a ubiquitous process in a biomolecule that a local perturbation of the structure, chemical or/and mechanical, induces a global conformational change of the molecule. The conformational change so induced affects the activity which biomolecules play in a variety of life phenomena \cite{Hirata_ref01,Hirata_ref02}.

One of such processes is seen in the biosynthesis in a bacterium in which a structural change of the retinal moiety bound at a residue of bacteriorhodopsin from ``\textit{cis}'' to ``\textit{trans}'' triggers a conformational change in the host molecule, which in turn induces successive chemical reactions including the charge transfer at the photoactive center \cite{Hirata_ref01}.

Another example of such conformational changes induced by a local chemical-perturbation is seen in the mutation or amino acid substitution of a protein, that induces a change in the global structure, which in turn causes acquisition or loss in activity of the biomolecule. Since the microorganism uses the mutation to survive against a drug, it is important to find the conformational change of target protein, induced by the amino acid substitution, as soon as possible \cite{Hirata_ref01,Hirata_ref03}.

The purpose of the present paper is to provide a theoretical framework to predict the global conformational change of a biomolecule, induced by a chemical or/and mechanical perturbation introduced at a moiety of the molecule, by means of the generalized Langevin theory (GLT) of the structural fluctuation of a biomolecule, or the Kim--Hirata theory, combined with the statistical mechanics theory of molecular liquids, or RISM/3D-RISM theory \cite{Hirata_ref04,Hirata_ref05}.

In 2013, Kim and Hirata  published a paper \cite{Hirata_ref04}, based on the generalized Langevin theory, that concerns a theoretical characterization of the structural fluctuation of a protein in solution. The most important conclusion the authors extracted from the theoretical results is that the force to restore the equilibrium conformation is proportional to the displacement vector, or fluctuation, of atoms in protein, just akin to a harmonic oscillator. The result is equivalent to say that the probability distribution of the conformational fluctuation is Gaussian. Therefore, the conformational fluctuation of protein in solution can be identified as a composite of many Gaussian-distributions with different variances, which are hierarchically ordered~\cite{Hirata_ref04,Hirata_ref05}.

A direct experimental evidence that the structural fluctuation of protein is Gaussian can be seen in the so-called Guinier plot of protein, in which the intensity of the small angle X-ray scattering (SAXS) is plotted against the square of wave number, $Q^{2}$ \cite{Hirata_ref06}. As it has been observed by Kataoka et al., the plot shows a linear behavior for a variety of conformations of protein, including native, denatured, molten globule, and so on, in the low wave vector region \cite{Hirata_ref07}. The behavior indicates unambiguously that the distribution of the conformational fluctuation of protein is Gaussian no matter what the structure is. Of course, the plot deviates from the straight line as the wave number increases, but it just reflects the hierarchical ordering of the modes, or the variance of a Gaussian distribution, which becomes less as the wave number $Q$ becomes greater \cite{Hirata_ref05,Hirata_ref07}.

The Gaussian behavior of the structural fluctuation of protein in solution is also verified by the molecular-dynamics simulation combined with the RISM/3D-RISM method, carried out by Chong and Ham \cite{Hirata_ref08,Hirata_ref09}. In the study, the authors  calculated the solvation free energy by means of the RISM/3D-RISM theory for each snapshot of the molecular dynamics trajectory, and found that the free energy including both the direct interactions among atoms in protein and the solvation free energy forms the Gaussian distribution, while the direct interaction by itself does not produce such distribution \cite{Hirata_ref08}.

It is a rational strategy in the statistical mechanics theory to apply the linear response theory, developed by Kubo, to such a system in which the fluctuational response is proportional to a perturbation \cite{Hirata_ref05,Hirata_ref10,Hirata_ref11}. In the same JCP paper in 2013, Kim and Hirata derived an equation of the linear response theory for the conformational change of a protein in water induced by a perturbation. The equation is in accord with that derived earlier by Ikeguchi et al. through an alternative route \cite{Hirata_ref04,Hirata_ref12}. The theory was applied to formulate the conformational change of protein induced by a thermodynamic perturbation such as pressure \cite{Hirata_ref13}.

In the present study, the author employs the linear response theory for the conformational change of a biomolecules, induced by a mechanical/chemical perturbation introduced at a moiety of the molecule. In what follows, the Kim--Hirata theory including the linear response theory is briefly reviewed, and is applied to the structural transition induced by a local conformational change in biomolecules.

\section{Theory}

\subsection{Brief review of the Kim-Hirata theory}

The Kim--Hirata theory \cite{Hirata_ref04} begins with the Liouville equation which describes the time evolution of dynamic variables ${\rm {\bf A}}(t)$ in the phase space, 
\begin{equation}
\label{Hirata_eq1}
\frac{\mathrm{d}{\rm {\bf A}}(t)}{\mathrm{d}t}=\mathrm{i}L{\rm {\bf A}}(t).
\end{equation}
In the equation, the vector ${\rm {\bf A}}(t)$ is defined by,
\begin{equation}
\label{Hirata_eq2}
{\rm {\bf A}}(t)\equiv \left( {{\begin{array}{*{20}c}
 {\Delta {\rm {\bf R}}_\alpha (t)} \\
 {{\rm {\bf P}}_\alpha (t)} \\
 {\delta \rho _a ({\rm {\bf r}},t)} \\
 {{\rm {\bf J}}_a ({\rm {\bf r}},t)} \\
\end{array} }} \right),
\end{equation}
where the Greek subscript $\alpha$ and the Roman subscript $a$ denote atoms in 
protein and solvent molecules, respectively. The variables $\Delta {\rm {\bf 
R}}_\alpha (t)$ and ${\rm {\bf P}}_\alpha (t)$ represent the structural 
fluctuation of protein, and its conjugate momentum, while $\delta \rho _a 
({\rm {\bf r}},t)$, and ${\rm {\bf J}}_a ({\rm {\bf r}},t)$ are the density 
fluctuation of solvent around protein and its momentum or the flux, defined 
by, 
\begin{gather}
\label{Hirata_eq3}
\Delta {\rm {\bf R}}_\alpha (t)\equiv {\rm {\bf R}}_\alpha (t)-\left\langle 
{{\rm {\bf R}}_\alpha } \right\rangle,
\quad
{\rm {\bf P}}_\alpha (t)\equiv M_\alpha \frac{\mathrm{d}\Delta {\rm {\bf R}}_\alpha 
}{\mathrm{d}t},
\\
\label{Hirata_eq4}
\delta \rho _a ({\rm {\bf r}},t)\equiv \sum\limits_i {\delta ({\rm {\bf 
r}}-{\rm {\bf r}}_i^a } (t))-\left\langle {\rho _a } \right\rangle ,
\quad
{\rm {\bf J}}_a ({\rm {\bf r}},t)\equiv \sum\limits_i {{\rm {\bf p}}_i^a } 
\delta ({\rm {\bf r}}-{\rm {\bf r}}_i^a ),
\end{gather}
where $\left\langle \cdots \right\rangle$ denote an ensemble average of the 
variables.

Following the recipe of the generalized Langevin theory (GLT), Kim and 
Hirata  projected all the mechanical variables in the phase space onto 
${\rm {\bf A}}(t)$ to derive essentially two GLEs for the time evolution of the 
dynamic variables, one for the dynamics of a biomolecule, the other for that 
of solvent \cite{Hirata_ref04,Hirata_ref05,Hirata_ref11,Hirata_ref14}. Here, we just focus on that relevant to the structural fluctuation of a solute molecule, which leads,
\begin{equation}
\label{Hirata_eq5}
M_\alpha \frac{\rd^2\Delta {\rm {\bf R}}_\alpha (t)}{\rd t^2}=-\sum\limits_\beta 
{A_{\alpha \beta } } \Delta {\rm {\bf R}}_\beta (t)-\int_0^t {\mathrm{d}s\sum\limits_ 
{\Gamma _{\alpha \beta } } } (t-s)\cdot \frac{\mathrm{d}\Delta {\rm {\bf R}}_\alpha 
(s)}{\mathrm{d}s}+W_\alpha (t),
\end{equation}
where the second and third terms in the right-hand-side represent the 
frictional force exerted by solvent and the random force due to the thermal 
agitation, which are related with each other by the fluctuation dissipation 
theorem. (Here, details of the expressions concerning the two terms are 
entirely skipped.) 

It is the first term on which we focus in the present paper, which looks like that of a harmonic oscillator: the restoring force is proportional to the displacement of atoms from their equilibrium positions, or to the structural fluctuation. In this respect, the equation is formally equivalent to that of a dumped harmonic oscillator in a viscus fluid. By neglecting the second and third terms of equation~(\ref{Hirata_eq5}), one finds an equation analogous to stationary dynamics of a harmonic oscillator, 
\begin{equation}
\label{Hirata_eq6}
M_\alpha \frac{\mathrm{d}^2\Delta {\rm {\bf R}}_\alpha (t)}{\mathrm{d}t^2}=-\sum\limits_\beta 
{A_{\alpha \beta } } \Delta {\rm {\bf R}}_\beta (t).
\end{equation}
In the equation, the characteristic or intrinsic frequency $A_{\alpha \beta}$ is related to ($\alpha ,\beta)$-element of the inverse of matrix \textbf{L} by 
\begin{equation}
\label{Hirata_eq7}
A_{\alpha \beta } =k_B T\left( {{\rm {\bf L}}^{-1}} \right)_{\alpha \beta }, 
\end{equation}
where \textbf{L} is the variance-covariance matrix of the structural 
fluctuation of the biomolecule, defined as,
\begin{equation}
\label{Hirata_eq8}
{\rm {\bf L}}\equiv \left\langle {\Delta {\rm {\bf R}}\Delta {\rm {\bf R}}} 
\right\rangle.
\end{equation}
The form of equation~(\ref{Hirata_eq6}) indicates that the energy surface to originate the restoring force is quadratic in the displacement vector or fluctuation, and the probability distribution of the fluctuation is Gaussian, the variance-covariance matrix of which is \textbf{L} defined by equation~(\ref{Hirata_eq8}). 

At this point, some readers may raise the following questions. Why  the free energy surface of the protein in water can possibly be quadratic? What is the capability of probability distribution of the structural fluctuation to become Gaussian? The quick answer to the question is: because it is a consequence of the central limiting theorem \cite{Hirata_ref05,Hirata_ref15,Hirata_ref16}. Of course, the potential energy surface of protein in water itself is never quadratic. As is seen in any computer-program of the molecular dynamics simulation, the interactions among atoms in protein as well as those with water molecules involve non-harmonic interactions, including the Lennard-Jones as well as Coulomb interactions. For such systems, the potential energy surface becomes strictly harmonic only when the system is cooled down to the global minimum. That is the essential requirement for the normal mode analysis (NMA) carried out earlier by several authors \cite{Hirata_ref17,Hirata_ref18}.

On the other hand, the protein structure in the thermal equilibrium is in the minimum of the free energy surface by definition, that consists of the interaction energy among atoms in protein and the solvation free energy, that is,
\begin{equation}
\label{Hirata_eq9}
F\left( {\left\{ {\rm {\bf R}} \right\}} \right)=U\left( {\left\{ {\rm {\bf 
R}} \right\}} \right)+\Delta \mu \left( {\left\{ {\rm {\bf R}} \right\}} 
\right),
\end{equation}
where $\left\{ {\rm {\bf R}} \right\}$ represents a set of coordinates of atoms in a biomolecule, and $U$ and $\Delta \mu$ denote the intramolecular interaction energy of the biomolecule and the solvation free energy, respectively \cite{Hirata_ref05}. Although it is not explicitly expressed, the quantity is a function of the solvent coordinates, the degrees of freedom of which are infinitely large. It is well regarded in the statistical mechanics that the fluctuation of such a system consisting of an infinite degrees of freedom strictly satisfies the central limiting theorem, and that the probability distribution of the fluctuation becomes Gaussian \cite{Hirata_ref05,Hirata_ref15,Hirata_ref16}
\begin{equation}
\label{Hirata_eq10}
w_{\text{conf}} \left( {\left\{ {\Delta {\rm {\bf R}}} \right\}} \right)=\sqrt 
{\frac{A}{\left( {2\piup } \right)^{3N}}} \exp \left[ 
{-\frac{1}{2}\sum\limits_\alpha {\sum\limits_\beta {A_{\alpha \beta } \Delta 
{\rm {\bf R}}_\alpha \Delta {\rm {\bf R}}_\beta } } } \right].
\end{equation}
Based on the logical entailment, Kim and Hirata  proposed an \textit{ansatz} that plays a crucial role for further developing the theory \cite{Hirata_ref04}. The ansatz is to equate the force constant $A_{\alpha \beta }$ of the restoring force acting on protein atoms, or the inverse of the variance-covariance matrix, to the second derivative of the free energy surface of a protein molecule in solution. That is,
\begin{equation}
\label{Hirata_eq11}
A_{\alpha \beta } =\frac{\partial ^2F\left( {\left\{ {\rm {\bf R}} \right\}} 
\right)}{\partial \Delta {\rm {\bf R}}_\alpha \partial \Delta {\rm {\bf 
R}}_\beta }.
\end{equation}
Since it is possible to calculate the solvation free energy $F\left({\left\{ {\Delta {\rm {\bf R}}} \right\}} \right)$ by means of the RISM/3D-RISM theory, the ansatz makes feasible the calculation of the force constant in solution. The ansatz has a mathematical isomorphism with the ordinary force constant $k_{\alpha \beta}$ in the harmonic oscillator, which is defined by,
\begin{equation}
\label{Hirata_eq12}
k_{\alpha \beta } =\frac{\partial ^2U\left( {\left\{ {\rm {\bf R}} \right\}} 
\right)}{\partial \Delta {\rm {\bf R}}_\alpha \partial \Delta {\rm {\bf 
R}}_\beta },
\end{equation}
where $U\left( {\left\{ {\rm {\bf R}} \right\}} \right)$ is the interaction energy among atoms in the molecule. 

The expression (\ref{Hirata_eq11}) for the force constant is quite useful to construct a linear response theory to describe conformational changes of a biomolecule, induced by some perturbation applied to the system. 

The equation (\ref{Hirata_eq11}) suggests that the free energy of protein at equilibrium is expressed in the following form:
\begin{equation}
\label{Hirata_eq13}
F\left( {\left\{ {\rm {\bf R}} \right\}} 
\right)=\frac{1}{2}\sum\limits_{\alpha \beta } {\Delta {\rm {\bf R}}_\alpha 
A_{\alpha \beta } \Delta {\rm {\bf R}}_\beta }.
\end{equation}
Let us apply a perturbation to the system, 
\begin{equation}
\label{Hirata_eq14}
F\left( {\left\{ {\rm {\bf R}} \right\}} 
\right)=\frac{1}{2}\sum\limits_{\alpha \beta } {\Delta {\rm {\bf R}}_\alpha 
A_{\alpha \beta } \Delta {\rm {\bf R}}_\beta } +\sum\limits_\alpha {\Delta 
{\rm {\bf R}}_\alpha \cdot {\rm {\bf f}}_\alpha }, 
\end{equation}
where ${\rm {\bf f}}_\alpha$ is the perturbation acting on the atom $\alpha$ of the molecule.The conformational change induced by the perturbation can be derived by the variational principle,
\begin{equation}
\label{Hirata_eq15}
\frac{\partial F\left( {\left\{ {\rm {\bf R}} \right\}} \right)}{\partial 
\Delta {\rm {\bf R}}_\beta }=0,
\end{equation}
which leads
\begin{equation}
\label{Hirata_eq16}
\left\langle {\Delta {\rm {\bf R}}_\alpha } \right\rangle _1 =\frac{1}{k_B 
T}\sum\limits_\beta {\left\langle {\Delta {\rm {\bf R}}_\alpha \Delta {\rm 
{\bf R}}_\beta } \right\rangle _0 \cdot {\rm {\bf f}}_\beta } .
\end{equation}

\subsection{Structural transition induced by a local conformational change in biomolecules}

The present section is devoted to the formulation of the theory to describe the structural change of a biomolecule induced by a local conformational change, such as the photo-excitation of a chromophore, and the substitution of an amino acid, based on the linear response theory described in the preceding section. For that purpose, the potential energy $U$ of the biomolecule in equation~(\ref{Hirata_eq9}) is decomposed into the three contributions as follows,
\begin{equation}
\label{Hirata_eq17}
U\left( {\left\{ {\rm {\bf R}} \right\}} \right)=U_r \left( {\left\{ {\rm 
{\bf R}} \right\}_r } \right)+U_m \left( {\left\{ {\rm {\bf R}} \right\}_m } 
\right)+U_{rm} \left( {\left\{ {\rm {\bf R}} \right\}_r ,\left\{ {\rm {\bf 
R}} \right\}_m } \right),
\end{equation}
where, $\left\{ {\rm {\bf R}} \right\}_m $ and $\left\{ {\rm {\bf R}} \right\}_r$ represent a set of coordinates of atoms in the moiety and that of reference protein without the moiety, respectively, $U_m$ and $U_r$ denote the potential energy of the respective portion of the protein, and $U_{rm}$ denotes interactions of atoms between the two portions. 

Now, we make a thought experiment in which only the moiety portion of the entire biomolecule is replaced by a new one. The difference in the potential energy before and after the replacement may be written as,
\begin{equation}
\label{Hirata_eq18}
\Delta U\left( {\left\{ {\rm {\bf R}} \right\}} \right)=\Delta U_m \left( 
{\left\{ {\rm {\bf R}} \right\}_m } \right)+\Delta U_{rm} \left( {\left\{ 
{\rm {\bf R}} \right\}_r ,\left\{ {\rm {\bf R}} \right\}_m } \right),
\end{equation}
where $\Delta U_m$ and $\Delta U_{rm}$ are the change in potential energy among atoms in the moiety and that between the moiety and the reference protein, respectively. 

The expression for the perturbation can be obtained by substituting $\Delta U$ into equation~(\ref{Hirata_eq16}) as, 
\begin{equation}
\label{Hirata_eq19}
{\rm {\bf f}}_\beta =-\frac{\partial \Delta U}{\partial {\rm {\bf R}}_\beta 
}=-\frac{\partial \Delta U_{rm} \left( {\left\{ {\rm {\bf R}} \right\}_r 
,\left\{ {\rm {\bf R}} \right\}_m } \right)}{\partial {\rm {\bf R}}_\beta 
},
\end{equation}
where ${\rm {\bf R}}_\beta $ denotes the coordinate of an atom in the reference system, and $\Delta U_{rm} $ is the difference between the interaction energy between atoms in the reference protein and those in the moiety before and after the moiety is modified. It should be noted that the derivative of $U_{m}$ disappeared, because it is irrelevant to the coordinate of the atom $\beta $ in the reference protein. 

By substituting equation~(\ref{Hirata_eq19}) into equation~(\ref{Hirata_eq16}), one finds,
\begin{equation}
\label{Hirata_eq20}
\left\langle {\Delta {\rm {\bf R}}_\alpha } \right\rangle _1 =\frac{1}{k_B 
T}\sum\limits_\beta {\left\langle {\Delta {\rm {\bf R}}_\alpha \Delta {\rm 
{\bf R}}_\beta } \right\rangle _0 \cdot \left( {-\frac{\partial \Delta 
U_{rm} \left( {\left\{ {\rm {\bf R}} \right\}_r ,\left\{ {\rm {\bf R}} 
\right\}_m } \right)}{\partial {\rm {\bf R}}_\beta }} \right)},
\end{equation}
in which 
$\left\langle{\Delta{\rm{\bf R}}_\alpha \Delta{\rm{\bf R}}_\beta}\right\rangle_0$ 
is the variance-covariance matrix of the reference system, that is, the protein without the moiety. The linear response expressions, equation~(\ref{Hirata_eq20}), is interpreted as follows. \textit{The force exerted by atoms in the moiety induces the displacement in atom }$\beta $\textit{ of protein, which propagates through the variance-covariance matrix} 
$\left\langle{\Delta{\rm{\bf R}}_\alpha \Delta{\rm{\bf R}}_\beta}\right\rangle_0$ 
\textit{to cause a global conformational change of the molecule,}
$\left\langle{\Delta{\rm{\bf R}}_\alpha}\right\rangle_1$. 

\begin{figure}[!t]
\centerline{\includegraphics[width=0.7\textwidth]{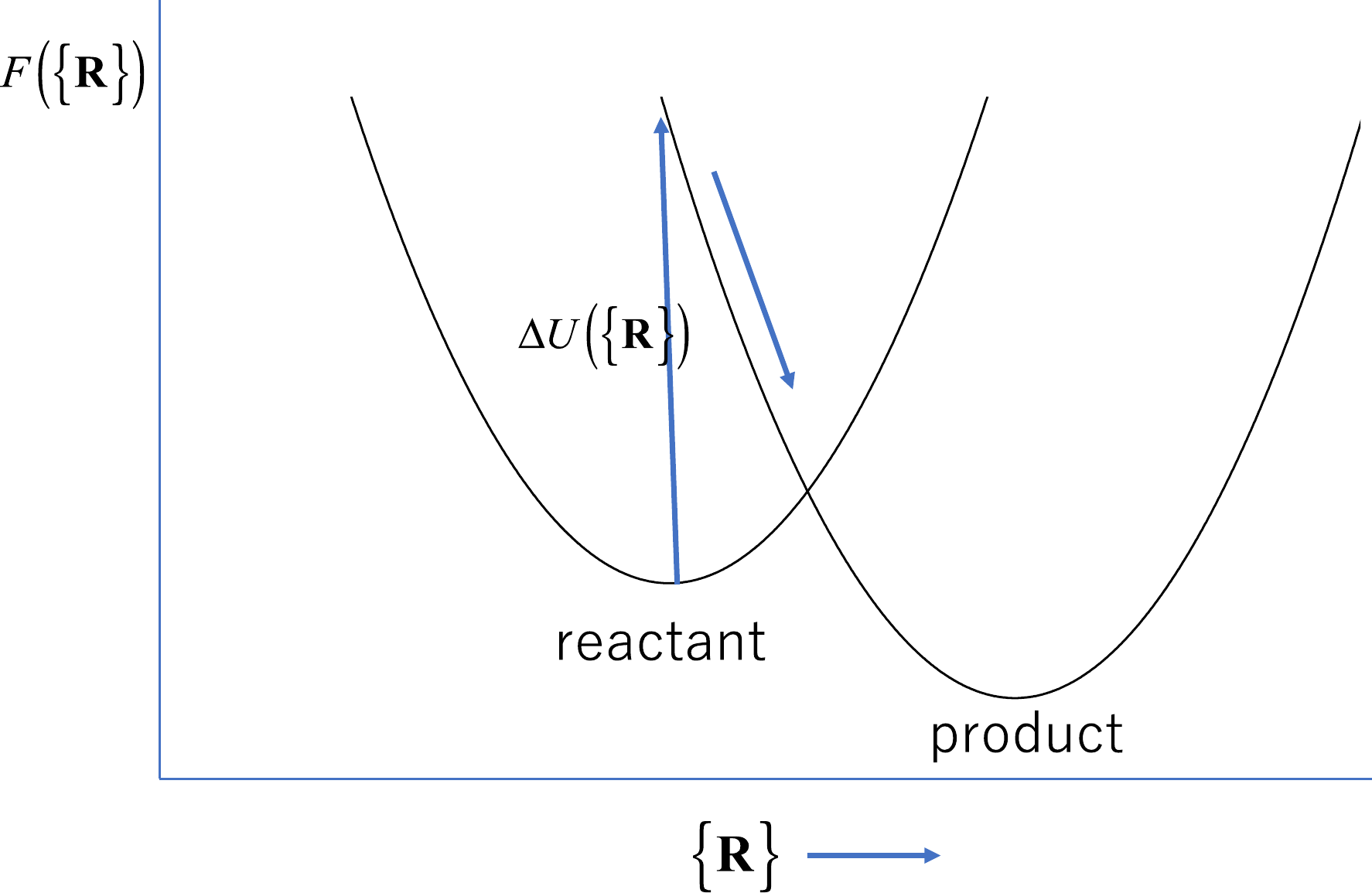}} 
\caption{(Colour online) Schematic picture of the reaction: the quadratic curves depict the free energy surface of reactant and product, and 
$\Delta U\left( {\left\{ {\rm {\bf R}} \right\}} \right)$
represents the perturbation defined by equation~(\ref{Hirata_eq18}).}
\label{Hirata_fig1} 
\end{figure}

The schematic picture that illustrates the reaction is presented in figure~\ref{Hirata_fig1}, in which the vertical excitation due to the perturbation, $\Delta U\left({\left\{ {\rm {\bf R}} \right\}} \right)$, induces the successive chemical reaction.

A question may be raised with respect to the applicability of equation~(\ref{Hirata_eq20}) to a process of mechanical and/or chemical perturbation. The question is as follows. The response to the perturbation may not necessarily be linear if the perturbation is too large, and the protein may lose the native conformation. In fact, there are many such cases taking place in nature, in which the protein loses its native conformation due to a chemical or mechanical perturbation. However, such cases are not of interest for us, because in such a case the perturbation causes the protein to lose its activity. Only proteins which may survive in nature are those that have some biological activity. In other words, we are interested in native protein, the structural fluctuation of which is linear against a perturbation. 

There is another concern with respect to the computational protocol to implement the theory to an actual process of structural change of protein. It may  largely depend on the computational procedure. Actual process to implement the theory can be treated by an idea proposed by Hirata and Akasaka in the context of the conformational change of protein induced by a thermodynamic perturbation \cite{Hirata_ref13}. The idea is to use the analytical continuation to keep the perturbation and response within the linear regime.

Let us divide the entire mechanical/chemical perturbation into $N$ steps, each step of which can be described by the linear response theory, or equation~(\ref{Hirata_eq20}). Then, the entire change of the conformation after $N$-steps may be expressed by, 
\begin{equation}
\label{Hirata_eq21}
\left\langle {\Delta {\rm {\bf R}}_\alpha } \right\rangle =\frac{1}{k_B 
T}\sum\limits_{j=1}^N {\sum\limits_\beta {\left\langle {\Delta {\rm {\bf 
R}}_\alpha \Delta {\rm {\bf R}}_\beta } \right\rangle _j \cdot \left( 
{-\frac{\partial \Delta U_{rm}^{(j)} \left( {\left\{ {\rm {\bf R}} 
\right\}_r ,\left\{ {\rm {\bf R}} \right\}_m } \right)}{\partial {\rm {\bf 
R}}_\beta }} \right)} } .
\end{equation}
For an implementation of the theory to an actual biophysical problem, a careful choice of $N$, or scheduling, may be required.

\section{Discussions}

The present section is devoted to discuss the feasibility of calculation to realize the theory derived in the preceding section, and to provide a few examples to which the theory may be applied.
In equation~(\ref{Hirata_eq19}), $\Delta U_{rm}$ and its derivative with respect to the atomic coordinate of a biomolecule can be calculated analytically by means of the molecular mechanics using the simulation programs such as AMBER \cite{Hirata_ref20}. Thus, it is a trivial problem for the current state of the art.

The crucial part of the calculation concerns the variance-covariance matrix\linebreak$\left\langle{\Delta{\rm{\bf R}}_\alpha \Delta{\rm{\bf R}}_\beta}\right\rangle_0$. 
The calculation is non-trivial, because it involves an ensemble average over the configuration space of solvent molecules, the number of which is $\sim 10^{23}$ or the thermodynamic limit. As is clarified in the introduction, such a limit is the requirement for the central limiting theorem as well as the linear response theory, on which the present treatment is based. It will be quite evident that the molecular dynamic simulation never meets the requirement, no matter what the hardware or/and the computational algorithm is.

It is the statistical mechanics of molecular liquids, or RISM/3D-RISM, that makes the calculation of 
$\left\langle{\Delta{\rm{\bf R}}_\alpha \Delta{\rm{\bf R}}_\beta}\right\rangle_0$
feasible, with the \textit{ansatz} expressed by equation~(\ref{Hirata_eq11}). According to the ansatz, the variance-covariance matrix is the inverse of the Hessian matrix, which is the second derivative of the free energy surface 
$F\left( {\left\{ {\rm {\bf R}} \right\}} \right)$, 
defined by equation~(\ref{Hirata_eq9}), with respect to the atomic coordinates of protein. It has been demonstrated recently by Sugita and Hirata \cite{Hirata_ref21} that the second derivative can be calculated along the numerical solution of the RISM/3D-RISM equations with the procedure proposed earlier by Yu and Karplus \cite{Hirata_ref22}. The spectrum of small wave-number regions of alanine dipeptide, calculated from the Hessian, showed a reasonable agreement with the results of the RIKES spectrum \cite{Hirata_ref23}. Applying the method to a real protein may require much larger computational time, but it is just the matter of a hardware, not the matter which touches a basic principle of nature, or the central limiting theorem.
Examples of possible applications of the theory proposed in the previous sections to a biophysical and/or biochemical process are discussed in what follows.

One of the applications  concerns the drug discovery. It is well documented that a crucial step for designing  drug compounds, either by a wet chemistry or in silico, is not only to identify a target protein but also to find its molecular structure. The current state of the art in the business is to determine the structure by means of the X-ray crystallography, or the two-dimensional NMR, both of which are quite laborious as well as time consuming. For example, in case of the X-ray crystallography, one has to crystallize the molecule first to be able to get the diffraction pattern, from which one extracts the structure. In case there are multiple candidate-proteins for a drug target, the process becomes formidable some time. If the mutation of the micro-organism is very quick, such efforts become entirely hopeless, as was demonstrated unambiguously by the latest pandemic of COVID-19. The theory proposed here can be applied to such a problem to find the structure of a target protein, which is derived from the native protein by an amino acid substitution. We still need to identify which amino acid is substituted by which. However, such an identification would be neither very difficult nor time-consuming by means of the current state of the art, for example, by the NMR spectroscopy.

Another example of the application is the mutagenesis to improve the activity of an enzyme, or to add a new activity to the biocatalyst \cite{Hirata_ref01}. In that case, too, it is important to confirm that the native conformation of the enzyme is intact after substituting an amino acid by a new one. If the biomolecule is denatured by the mutation, it will lose or reduce the activity as a biocatalyst. Since the substitution of amino acid is made currently in trial and error basis by means of a wet chemistry, the process becomes quite laborious as well as time consuming. Therefore, it is desirable to predict the structure of the biomolecule after each trial of an amino acid substitution employing the method proposed in the present paper.

\section{Concluding remarks and perspective}

Based on the Kim--Hirata theory for the structural fluctuation of a biomolecule in aqueous solution, the author has proposed a new theoretical method to predict the conformation of the biomolecule, a moiety of which is modified chemically or/and mechanically by such a process as photo-excitation or amino acid substitution. The method is expected to be applied to the medical as well as biomimetic processes in industry, in order to find the conformation of biomolecules after chemical or/and mechanical modifications. It is a key to the  successful application of the RISM/3D-RISM method  in order to calculate the free energy surface of the protein, including solvation free energy, and its second derivative with respect to the atomic coordinate of the biomolecule.

The benefit of using the RISM/3D-RISM is not limited to the application described in the present paper. Using the 3D-RISM theory in a series of chemical processes may accelerate an industrial innovation, for example, the discovery of a new drug. Let us mention such two examples in which the use of RISM/3D-RISM theory may accelerate the industrial processes.

One of such processes concerns the drug discovery. A drug discovery consists of many steps; finding the target protein, screening the candidate compounds for the target protein, synthesizing the compounds, solubility tests, clinical tests, and so forth. Among those steps, it is the compound screening in which the in-silico approach may play a crucial role. The compound screening is to find a compound among many candidate compounds, which has the highest binding affinity to the target protein, malfunctioning of which is fatal for the host micro-organism. The RISM/3D-RISM method has been applied to many target proteins and the candidate compounds to predict the binding affinity successfully. However, the applications have been limited so far to those in which the structural information of the target protein is provided. Such cases in which the conformation is unknown has been entirely ``out of scope'' for the method. The method proposed in the present paper may be applied to such cases to determine the conformation of biomolecules created by mutations. Then, the screening of drug compounds may be dramatically accelerated, because the wet chemistry to determine the structure of mutant protein is replaced also by the in-silico process.

The other process, to which the present theory may be applied, is the rate of chemical reaction in biomolecular solutions. An example of such reactions is the photochemical reaction, in which the reaction induced at a chromophore, such as a \textit{cis-trans} isomerization of retinal, triggers a successive conformational change in the entire protein. The initial photochemical reaction is so fast that the conformational change of the entire protein becomes the rate-determining step. Recently, the author has proposed an Arrhenius-type theory of the reaction rate, in which the activation barrier is defined by the crossing point of two parabolas representing the free energy surfaces of the reactant and product along the reaction coordinate \cite{Hirata_ref24}. It is supposed in the theory that the equilibrium conformations of the both reactant and product states are provided. Then, all the quantities required to calculate the reaction rate, the equilibrium free energy of the reactant and product states as well as the curvature of the free energy surfaces, can be calculated by means of the RISM/3D-RISM method. It may be more productive if one can predict the structure of the product state \textit{in silico} by method proposed in the present paper, since it will help to skip a wet chemistry to determine the structure of the product state.

\section{Acknowledgement}

The author is a professor emeritus of Institute for Molecular Science, and a recipient of Toyota Riken Fellow (2016--2019). He is grateful to M.~Terazima and Y.~Watanabe for the useful comments to motivate the present research.

%
%

\ukrainianpart

\title{Структурний перехід під впливом локального хімічного/механічного збурення в біомолекулах} 
\author{Ф. Хірата}
\address{Національний інститут природничих наук, Інститут молекулярних досліджень, Міодайджі, Окадзакі, Айті 444--8585, Японія}

\makeukrtitle

\begin{abstract} 
	\tolerance=3000%
	Структурний перехід під впливом локальної конформаційної перебудови у біомолекулах описано на основі узагальненої теорії Ланжевена для структурних флуктуацій молекули в розчині та теорії лінійного відгуку, розробленої Кімом і Хіратою в 2012 р. Хімічна/механічна перебудова, що відбувається у фрагменті біомолекул така як аміно-кислотне заміщення чи структурна перебудова хромофора при фотозбудженні, трактується як збурення, а решта протеїну --- як система відліку. Рівняння лінійного відгуку складається з двох частин: тієї, яку формує механічне/хімічне збурення, внесене фрагментом біомолекули, та іншої, що задається коваріаційною матрицею системи відліку і пов'язана з узагальненою сприйнятливістю. Це рівняння має прозорий фізичний зміст: сила, з якою діють атоми у фрагменті біомолекули, викликає зміщення атома протеїну, поширення якого відображається через коваріаційну матрицю, що приводить до глобальної конформаційної перебудови молекули.
	Запропоновано кілька прикладів можливого застосування теорії, в тому числі в промисловості.
	\keywords структурні фазові переходи, рівняння Ланжевена, теорія лінійного відгуку, біомолекули
	
\end{abstract}

\end{document}